\documentclass[aps,pre,twocolumn,superscriptaddress,showpacs,amssymb]{revtex4}
\usepackage{graphicx}
\usepackage{hyperref}
\usepackage{natbib}
\usepackage{amssymb}
\usepackage{color}
\usepackage{amsmath}
\usepackage{bm}
\begin{document}

\title{Suppression of quantum chaos in a quantum computer hardware}

\author{J. Lages}
\altaffiliation[Present address: ]{Laboratoire de Physique
Mol\'eculaire, UMR CNRS 6624, Universit\'e de Franche-Comt\'e - La
Bouloie, 25030 Besan\c con Cedex, France. E-mail:
jose.lages@univ-fcomte.fr}
\author{D. L. Shepelyansky}
\affiliation{Laboratoire de Physique Th\'eorique, UMR 5152 du CNRS,
Universit\'e Paul Sabatier, 31062 Toulouse Cedex 4, France}
%\date{\today}
\date{October 14, 2005}
%05.45.Mt Quantum chaos; semiclassical methods
%05.45.Pq Numerical simulations of chaotic systems
%03.65.Yz Decoherence; open systems; quantum statistical methods

\begin{abstract}
We present numerical and analytical studies
of a quantum computer proposed by the Yamamoto group in
Phys. Rev. Lett. {\bf 89}, 017901 (2002).
The stable and quantum chaos regimes in the quantum computer hardware
are identified as a function of magnetic field gradient
and dipole-dipole couplings between qubits on a square lattice.
It is shown that a strong magnetic field gradient
leads to suppression of quantum chaos.
\end{abstract}
\maketitle

\section{Introduction}
The fascinating idea of quantum computation stimulated numerous
experimental efforts for realization of qubits based on various
physical implementation of two-level quantum systems with
controllable interactions (see e.g. \cite{Nielsen}). Liquid-state
NMR implementation of quantum computation allowed to realize various
quantum algorithms with few qubits including the Shor factoring
algorithm \cite{Vandersypen1} and complex dynamics simulations
\cite{corybaker}. However, the liquid-state NMR scheme does not
allow individual addressing of qubits and thus cannot lead to a
large scale quantum computations \cite{schack}. The situation can be
significantly improved in the case of  solid-state NMR
implementation of quantum computation \cite{Nielsen}. Certain
features of this scheme appear also in the all-silicon quantum
computer recently proposed by the Yamamoto group
\cite{Yamamoto,Ladd1,Ladd2,Ladd3}. In this proposal the qubits are
spin-halves nuclei (e.g. isotopes ${^{29}}$Si) placed on a 2D
lattice on a surface of a crystalline solid matrix (e.g. of spin-0
$^{28}$Si nuclei). A magnetic field gradient is assumed to be
applied in the plane of the lattice to allow address qubits
individually. At present large gradients can be realized
experimentally \cite{Rugar,Rugar1} and according to
\cite{Yamamoto,Ladd3} thousands of qubits can be addressed. In
addition to qubit frequency gradient there are also dipole-dipole
couplings between qubits typical of liquid-state NMR
\cite{Slichter,Vandersypen}. These two elements, frequency gradient
and dipole-dipole couplings between qubits, also appear in other
proposals of quantum computers: e.g. for trapped polar molecules in
electric field with gradient \cite{DeMille} and trapped-ion spin
molecules with magnetic gradient \cite{Twamley}. Therefore the
investigation of generic properties of such systems is important for
future experimental implementations.

Indeed, it is known that the residual couplings between
qubits lead to emergence of quantum chaos
and melting of quantum computer hardware
\cite{georgeot00,flambaum,dls2001}.
It has been also shown \cite{benenti,frahm} that these static imperfections
give a rapid decrease of fidelity of quantum computations
and hence their analysis is important to preserve high
computation accuracy.
For a 1D chain of qubits it has been shown
that the introduction of frequency gradient
generally makes the quantum hardware more stable
against static inter-qubit couplings. This result
may have important implications for proposals similar to those as
in \cite{DeMille,Twamley}. However, a more generic 2D case
proposed in \cite{Yamamoto} still requires a special study.
Such a detailed study is the main aim of this paper.
To analyze the generic properties
the quantum computer hardware \cite{Yamamoto} we performed extensive
numerical simulations with up to 18 qubits. Our studies
show that a sufficiently strong field gradient
leads to a suppression of quantum chaos
and emergence of integrable regime
where the real eigenstates
become close to those of the ideal
quantum computer without imperfections.
On the basis of obtained results we determine
the border between this integrable regime and quantum chaos
region with ergodic eigenstates
for various system parameters.

The structure of the paper is the following:
in Section II we introduce typical physical parameters
of the system, the results of numerical simulations and
analytical estimates are presented in Section III
and the discussion and conclusions are given in
Section IV.

\begin{figure}[h]
\includegraphics[width=0.6\columnwidth]{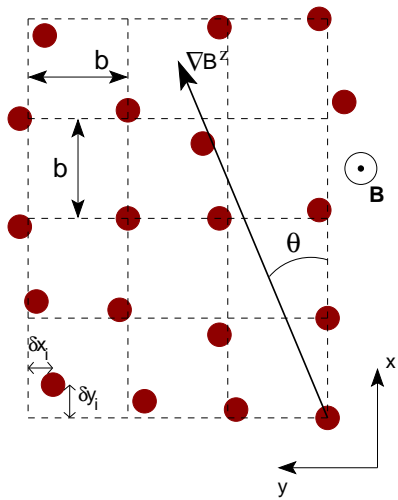}
\caption{\label{reseau}Disordered spin lattice: Local deviation from
the vertex position $(n_{x_i}b,n_{y_i}b)$ is given by $\delta
x_i=\delta_{x_i}b$ and $\delta y_i=\delta_{y_i}b$ (see text). The
index $i$ labels the spins and runs from $1$ to $n_q=n_xn_y$. The
magnetic field gradient has angle $\theta$ with $x$-axis (shown by
arrow).}
\end{figure}

\section{Description of quantum computer model}

Following the Yamamoto group proposal \cite{Yamamoto}
we introduce here a mathematical (YQC) model
of a realistic quantum computer with field gradient
and dipole-dipole couplings between qubits.
For that
we consider a two-dimensional array of $n_q$ nuclear spins 1/2
embedded in a crystalline solid, \textit{e.g.} spin-1/2 $^{31}$P
nuclei in a GaAs/AlGaAs quantum well or spin-1/2 $^{29}$Si nuclei
in a matrix of spin-0 $^{28}$Si nuclei \cite{Yamamoto}. In order
to manipulate the nuclear spins using radio-frequency (RF) fields
each qubit should be distinguishable, \textit{i.e.} a specific Larmor
frequency has to be assigned to each of them. The Hamiltonian of
this array of nuclear spins is the following
\begin{equation}\label{eq1}
\begin{array}{lcl}
H&=&\sum_{i=1}^{n_q}
\left(\omega_0+\delta\omega_i\right)I_i^z\\
&+&\sum_{i<j}d_{ij}\left( 2I_i^zI_j^z -I_i^xI_j^x -I_i^yI_j^y
\right).
\end{array}
\end{equation}
This Hamiltonian is widely used in NMR studies \cite{Slichter}.
The operators $I_i^{x,y,z}= \prod_{j=1}^{i-1} \mathbf{1}_2 \otimes
\frac12\sigma^{x,y,z} \otimes \prod_{j=i+1}^{n_q} \mathbf{1}_2$
are the spin operators acting on the $i$th spin where
$\sigma^{x,y,z}$ are the usual Pauli matrices. The frequency
$\omega_0+\delta\omega_i$ is the Larmor frequency of the $i$th
spin. Considering the spin-1/2 nuclei as an ensemble of qubits,
$\omega_0+\delta\omega_i$ is the energy between the two states
$\left\arrowvert\uparrow\right\rangle,
\left\arrowvert\downarrow\right\rangle$ (or $\left\arrowvert
0\right\rangle, \left\arrowvert 1\right\rangle$) of a single
qubit. The double sum runs over all the nuclei on the xy-plane and
the term $d_{ij}=d/q^3_{ij}$ is the dipole coupling between the
spins $i$ and $j$. Here $q_{ij}=r_{ij}/b$ is the distance between two
spins $i$ and $j$ in units of the spin lattice constant $b$;
$d$ is the coupling between two nearby spins on the square lattice
($q_{ij}=1$). In the Faraday geometry, \textit{i.e.} the geometry
where the magnetic field is orthogonal to the plane of the
spin-1/2 nuclei ($\mathbf{B}=B^z\mathbf{\hat{z}}$), the coupling
constant $d$ is given by $d=\mu_0\gamma^2\hbar^2/8\pi b^3$ where
$\gamma$ is the gyromagnetic ratio of the considered nuclei.
A schematic view of the structure is shown in Fig.~1.
In the present
work we keep focus on the regime $\omega_0\gg\delta\omega_i,d$
where the single qubit energy $\omega_0$ is much bigger than the Larmor
frequency shifts $\delta\omega_i$ and bigger than the dipole
coupling $d_{ij}$ between the qubits. This corresponds to typical experimental
conditions discussed in \cite{Yamamoto,Ladd3}.

In (\ref{eq1}) a particular Larmor frequency shift
$\delta\omega_i$ is assigned to each nuclear spin. These Larmor
frequency shifts are needed to distinguish one nuclear spin from
the other, and can be seen as playing the same role as chemical
shifts in liquid NMR \cite{Slichter}. Here, these Larmor frequency
shifts originate from a magnetic field gradient
$\mathbf{g}=\boldsymbol{\nabla}B^z=g\mathbf{\hat{z}}$ lying on the
xy-plane. For the $i$th nuclear spin located at the position
$\mathbf{r}_i=n_{x_i}b\mathbf{\hat{x}} +n_{y_i}b\mathbf{\hat{y}}$
the Larmor frequency shift is
\begin{equation}\label{eq2}
\delta\omega_i=\frac{\gamma}{2\pi}
\mathbf{r}_{i}\cdot\boldsymbol{\nabla}B^z =
n_{x_i}\omega_g\cos\theta + n_{y_i}\omega_g\sin\theta
%\frac{\gamma}{2\pi}\left(
%x_ig\cos\theta
%+y_ig\sin\theta\right)
\end{equation}
where we define the characteristic frequency shift
$\omega_g=\frac{\gamma}{2\pi}bg$. The Larmor frequency spacing
between two nuclear adjacent spins along the $\hat{\mathbf{x}}$
(respectively the $\hat{\mathbf{y}}$) direction is
$\omega_{gx}=\omega_g\cos\theta$ (respectively
$\omega_{gy}=\omega_g\sin\theta$). In Tab. 1 typical physical
parameters are given for the case of $^{31}$P and $^{29}$Si
nuclei. For example, if we consider
a face-centered-cubic lattice of $^{31}$P nuclei with nearest
neighbor distance $b=3.9974\mathrm{\AA}$ then $d=154\mathrm{Hz}$
and the regime $\omega_g/d=10$ in our simulations
corresponds to $\omega_g=1540\mathrm{Hz}$ which is equivalent to a
magnetic field gradient $g=0.224\mathrm{T/\mu m}$. If we consider
now a lattice of $^{29}$Si nuclei with nearest neighbor distance
$b=1\mathrm{\AA}$ then $d=2.4\mathrm{kHz}$ and, for example, the
regime $\omega_g/d=10$ in our simulations corresponds to
$\omega_g=24\mathrm{kHz}$ which is equivalent to a magnetic field
gradient $g=28\mathrm{T/\mu m}$.

\begin{table}[h]\label{tab1}
\begin{tabular}{|l||r|r|r|r|r|}
\hline & $|\gamma/2\pi|$ & $b$ & $d$ & $\omega_g=10d$ & $g\sim
10d/b$ \\
\hline $^{31}$P  &  17.2 MHz/T& 4    \AA& 154  Hz& 1.54
kHz& 0.224 T/$\mu$m \\
\hline $^{29}$Si &  8.47 MHz/T& 1.9  \AA& 346
Hz& 3.46 kHz&  2.15 T/$\mu$m \\
\hline $^{29}$Si &  8.47 MHz/T& 1
\AA& 2374 Hz& 23.74 kHz& 28.0 T/$\mu$m \\ \hline
\end{tabular}
\caption{Typical physical parameters for different nuclei
corresponding to $\omega_g/d=10$ in our simulations.
}
\end{table}

For a case of square lattice we
consider the index $i$ labeling row by row each nucleus in the
array and thus running from 1 to $n_y$ (first row) then from
$n_y+1$ to $2n_y$ (second row) and this until $i=n_q=n_xn_y$.
The angle $\theta$ for the direction of
magnetic field gradient is measured in radians
and due to symmetry we assume $\theta \leq \pi/4$.
Experimentally, it is convenient to assign to the nuclei along the
path indexed by $i$ the Larmor frequencies in ascending order.
This ensures a total distinguishability of the nuclei. This
condition is realized if
$n_y\omega_g\sin\theta<\omega_g\cos\theta$. As the number of
nuclei measured in a quantum dot can be as large as
$n_q=n_xn_y\sim10^4$, this condition becomes $\theta<n_y^{-1}$.
Along the $\hat{\mathbf{x}}$ direction the Larmor frequency
spacing between nearest neighbors is then
$\omega_{gx}\simeq\omega_g\gg d$ and as a consequence the nuclei
along that direction are highly distinguishable. The dipole
coupling in (\ref{eq1}) is then negligible. Along the
$\hat{\mathbf{y}}$ direction the Larmor frequency spacing between
nearest neighbors is $\omega_{gy}\simeq\omega_g/n_y\ll\omega_g$,
the nuclei along that direction are weakly distinguishable. With
$n_y\sim 10^2$ and using typical physical parameters of Tab. 1, we
easily remark that along the $\hat{\mathbf{y}}$ direction the
dipole couplings $d_{ij}$ cannot be neglected since
$d\sim\omega_{gy}$.

In order to be more realistic we consider now the fact that the
spins cannot form an ideal lattice, \textit{i.e.} experimentally
it is not possible to place each nucleus exactly on a regular
rectangular lattice vertex. Consider the spin $i$ at the vicinity
of the lattice vertex $(n_{x_i},n_{y_i})$. We define the
deviations $\delta_{x_i}$ and $\delta_{y_i}$ which characterize
the deviation of the spin $i$ from the ideal vertex position
$(n_{x_i},n_{y_i})$. The position of the spin $i$ is then
$\mathbf{r}_i=\left(n_{x_i}+\delta_{x_i}\right)b\mathbf{\hat{x}}
+\left(n_{y_i}+\delta_{y_i}\right)b\mathbf{\hat{y}}$ (see Fig. 1).
We assume that the deviations $\delta_{x_i}$ and $\delta_{y_i}$ are
random and distributed in the interval
$\left[-\delta/2,\delta/2\right]$ where $\delta$ is determined by an
unavoidable experimental error in the nuclei positions. These
spatial deviations modify weakly the dipole couplings $d_{ij}$
between nuclear spins as it depends on the inverse of the third
power of the spin-spin distance $r_{ij}$. The main effect is the
modification of the Larmor frequency shifts and as a consequence
the Larmor frequency spacings between nuclear spins. For the $i$th
spin the Larmor frequency shift is now
\begin{equation}\label{eq3}
\delta\omega_i= \left(n_{x_i}+\delta_{x_i}\right)\omega_{gx} +
\left(n_{y_i}+\delta_{y_i}\right)\omega_{gy}.
\end{equation}
Thus the smallest Larmor frequency spacing between adjacent
nuclear spins is
\begin{equation}\label{eq4}
\begin{array}{lcl}
\omega_{gx}'&=&\min\left\{\delta\omega_{i+n_y}
-\delta\omega_i\right\}_{i=1,n_q-n_y}\\
&=&\left(1-\delta\right)\omega_{gx} -\omega_{gy}\delta
\end{array}
\end{equation}
along the $\hat{\mathbf{x}}$ direction and
\begin{equation}\label{eq5}
\begin{array}{lcl}
\omega_{gy}'&=&\min\left\{\delta\omega_{i+1}
-\delta\omega_i\right\}_{i=1,n_q-1}\\
&=&\left(1-\delta\right)\omega_{gy}-\omega_{gx}\delta
\end{array}
\end{equation}
along the $\hat{\mathbf{y}}$ direction. For $\theta\in[0,\pi/4]$
and for $\delta<0.5$, $\omega_{gx}'>0$ which ensures that along
the path $i$ from one row to another the Larmor frequency
increases. As we discussed before it is experimentally convenient
that $\delta\omega_{i+1}-\delta\omega_i>0$ for each couple of
adjacent Larmor frequency inside a row, \textit{i.e.} along the
$\hat{\mathbf{y}}$ direction. Using (\ref{eq5}) this condition
leads to the following inequality
$\tan\theta>\delta/\left(1-\delta\right)$. Summarizing, for a
large number $n_y$ of nuclei along the $\hat{\mathbf{y}}$
direction and small spatial imprecision $\delta\ll1$ total
distinguishability of the nuclei in the xy-plane will be achieved
if the following condition is fulfilled
\begin{equation}\label{eq6}
\delta<\theta<n_y^{-1}.
\end{equation}
Note that the disorder introduced in (\ref{eq3}) can also model the
spatial inhomogeneity of the two-dimensional magnetic field gradient
$\mathbf{g}$. Since the condition $\delta<n_y^{-1}$ has to be
realized the condition of total distinguishability (\ref{eq6})
imposes stringent experimental precision on the nuclei positions and
on the two-dimensional magnetic field gradient.

%the Larmor frequency spacings will be distributed in a window
%frequency of width
%$2\delta\omega_g\left(\cos\theta+\sin\theta\right)$.

In this paper we consider a typical experimental situation
when the
dipole couplings and the Larmor frequency shifts have
comparable values
$d \sim \omega_g$ but
$\omega_0\gg\omega_g,d$. In this regime, eigenstates of the
Hamiltonian (\ref{eq1}) can be ordered by spin sectors
$\left\langle I^z\right\rangle
=\left\langle\sum_iI_i^z\right\rangle$. Eigenstates with the same
$\left\langle I^z\right\rangle$ value form an energy band of width
$\delta\omega\simeq\omega_g\sqrt n_q$. Within this band
$n_{\left\langle I^z\right\rangle} =n_q!/(n_q/2+\left\langle
I^z\right\rangle)! (n_q/2-\left\langle I^z\right\rangle)!$
eigenstates are contained. Nearest bands, \textit{i.e.} bands with
a difference of $\pm1$ in $\left\langle I^z\right\rangle$, are
well separated in energy since their spacing is
$\omega_0\gg\delta\omega$. In this case, using spin flip operators
$I^{\pm}_j=I_j^x\pm iI_j^y$, the Hamiltonian
(\ref{eq1}) can be rewritten as
\begin{equation}\label{eq7}
H=H_{\mathrm{diag}}+H_{\mathrm{off-diag}}
\end{equation}
where
\begin{equation}\label{eq8}
H_{\mathrm{diag}}=\sum_{i=1}^{n_q}
\left(\omega_0+\delta\omega_i\right)I_i^z+2\sum_{i<j}d_{ij}I_i^zI_j^z.
\end{equation}
is the diagonal part of the Hamiltonian
and where
\begin{equation}\label{eq9}
H_{\mathrm{off-diag}}=-\frac12\sum_{i<j}d_{ij}\left(
I_i^+I_j^-+I_i^-I_j^+ \right)
\end{equation}
is the off-diagonal part. From (\ref{eq8}) we straightforwardly
remark that the off-diagonal part of the Hamiltonian (\ref{eq7}) is
block diagonal, \textit{i.e.} each block corresponds to a spin
sector $\left\langle I^z\right\rangle$. This means that only
noninteracting states with the same total spin value $\left\langle
I^z\right\rangle$ are directly coupled by the dipole interaction.
The diagonalization of any spin sector $\left\langle
I^z\right\rangle$ can be performed independently of the others.

The Hamiltonian (\ref{eq7}) determines the mathematical YQC model we
study in this paper. Since the quantum chaos first appears in the
middle of the spectrum \cite{georgeot00}, we study the properties of
the Hamiltonian (1) in the central band which is associated to the
$\left\langle I^z\right\rangle=0$ spin sector (respectively
$\left\langle I^z\right\rangle=\pm 1/2$ spin sector) for even
(respectively odd) number of spins. This part of the spectrum is
most sensitive to imperfections and emergence of quantum chaos first
starts here.

\section{Numerical results}

In order to characterize the emergence of quantum
chaos in the YQC model (\ref{eq7}) we use the level spacing
distribution $P(s)$ which clearly
marks the transition from
integrable to ergodic eigenstates (see e.g. \cite{Haake}
and Refs. therein).
Here, $s=(\epsilon_{i+1}-\epsilon_i)/\Delta$ is
the normalized level spacing
for energy levels
$\epsilon_i$ in a close vicinity of the middle of the central
band (we use 10\% of the spectral bandwidth $\delta\omega$);
$\Delta=\left\langle\epsilon_{i+1}-\epsilon_i\right\rangle$ is the
mean level spacing in this vicinity. The form of $P(s)$
is linked to the ergodic properties
of eigenstates: an integrable system
has $P(s)$ in
the form of the Poissonian distribution $P_{P}(s)=\exp(-s)$,
and $P(s)$ is given by the Wigner-Dyson distribution
of the random matrix theory
$P_{WD}(s)=\left(\pi/2\right)s\exp(-\pi s^2/4)$ if the eigenstates
are ergodic. Fig.~2 shows typical examples of the level
spacing distribution $P(s)$ for a disorder strength $\delta=0.1$
and for a magnetic field gradient angle $\theta=0.3\mathrm{rad}$.
Each distribution $P(s)$ in Fig.~2 have been calculated using
128700 energy levels and $N_d=10$ disorder configurations.  As
$\omega_g/d$ increases from $\omega_g/d=0$ to $\omega_g/d=10$ the
distribution $P(s)$ changes from the Wigner-Dyson distribution to the
Poissonian distribution. Thus the YQC hardware is in the
regime of quantum chaos for $\omega_g<\omega_g^c$
while for $\omega_g>\omega_g^c$ it is in the integrable regime.
According to the data of Fig.~2 the value of $\omega_g^c$
is located in the interval $4<\omega_g^c/d<7$.

\begin{figure}
\includegraphics[width=\columnwidth]{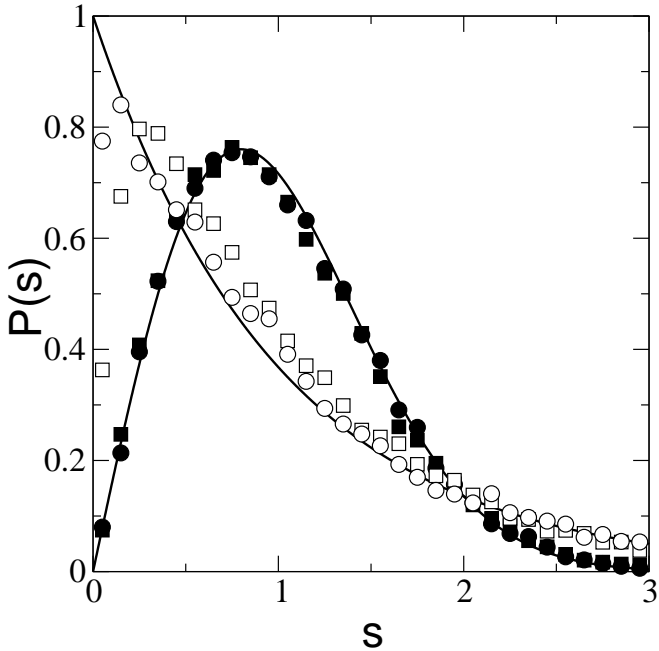}
\caption{The level spacing distribution $P(s)$ for a case with
disorder fluctuations amplitude $\delta=0.1$, for $\theta=0.3$, and
for $\omega_g/d=1$ ($\bullet$), 4 ($\blacksquare$), 7 ($\square$),
and 10 ($\circ$). The solid curves show the Poissonian and the
Wigner-Dyson distributions. The number of disorder realizations is
$N_d=10$. There are $n_q=16$ qubits placed on  the lattice of size
$4\times 4$.}
\end{figure}

\begin{figure}
\includegraphics[width=\columnwidth]{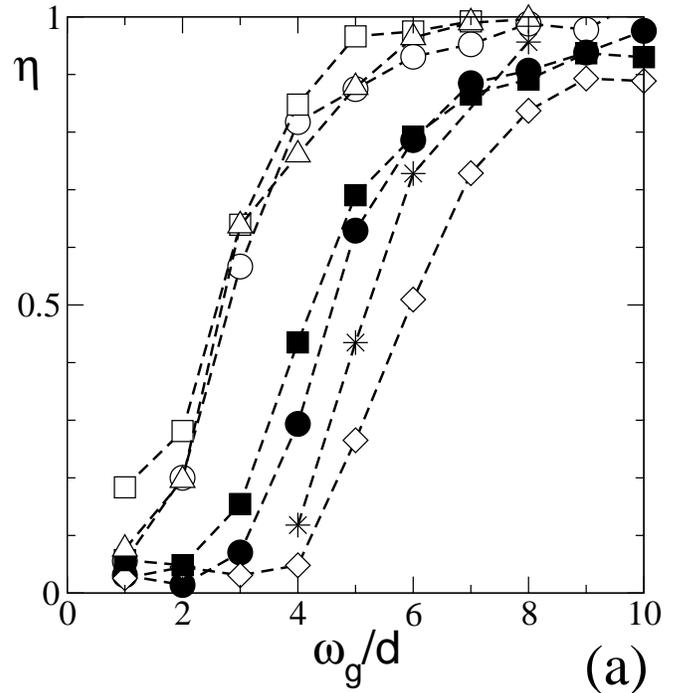}
\includegraphics[width=\columnwidth]{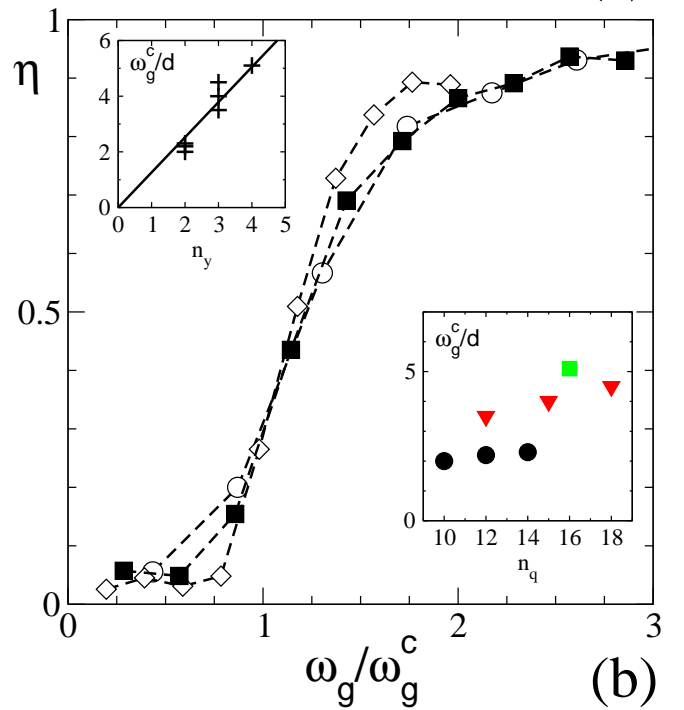}
\caption{(a) Parameter $\eta$ as a function of $\omega_g/d$ for
different lattice sizes  $5\times2$ ($\square$), $6\times2$
($\triangle$), $7\times2$ ($\circ$), $4\times3$ ($\blacksquare$),
$5\times3$ ($\bullet$), $6\times3$ ($\ast$), and $4\times4$
($\diamond$). (b) Parameter $\eta$ as a function of the rescaled
frequency shift $\omega_g/\omega_g^c$ where $\omega_g^c$ is defined
by $\eta(\omega_g^c/d)=0.3$. For clarity only some curves presented
in (a) are shown rescaled in (b). Upper inset: critical frequency
shift $\omega_g^c$ as a function of $n_y$, the straight line shows
the linear fit (see text). Lower inset: critical frequency shift as
a function of the total number of qubits $n_q=n_xn_y$, where circles
corresponds to $n_y=2$, triangles to $n_y=3$, and the square to
$n_y=4$. In all cases $\delta=0.1$ and $\theta=0.3$. The number of
disorder realization is $N_d=10$ ($N_d=4$ for the case $6\times3$).}
\end{figure}

To investigate the quantum chaos border $\omega_g^c$
in more detail
it is convenient to introduce the parameter \cite{Jacquod}
\begin{equation}\label{eq10}
\eta=\frac{\int_0^{s_0}\left[P(s)-P_{WD}(s)\right]ds}{\int_0^{s_0}
\left[P_P(s)-P_{WD}(s)\right]ds}
\end{equation}
where $s_0\simeq0.4729$ is the first intersection point between the
two distributions $P_P(s)$ and $P_{WD}(s)$. In this way $\eta$ tends
to $0$ if the system is chaotic ($P(s)\rightarrow P_{WD}(s)$) and
tends to $1$ if the system is integrable ($P(s)\rightarrow
P_{P}(s)$). This parameter $\eta$ is rather convenient for
investigation of the transition from integrability to quantum chaos
and it has been  used in various studies of many-body quantum
systems \cite{georgeot00,Jacquod,georgeotshard,benentiepjd}.

\begin{figure}
\includegraphics[width=\columnwidth]{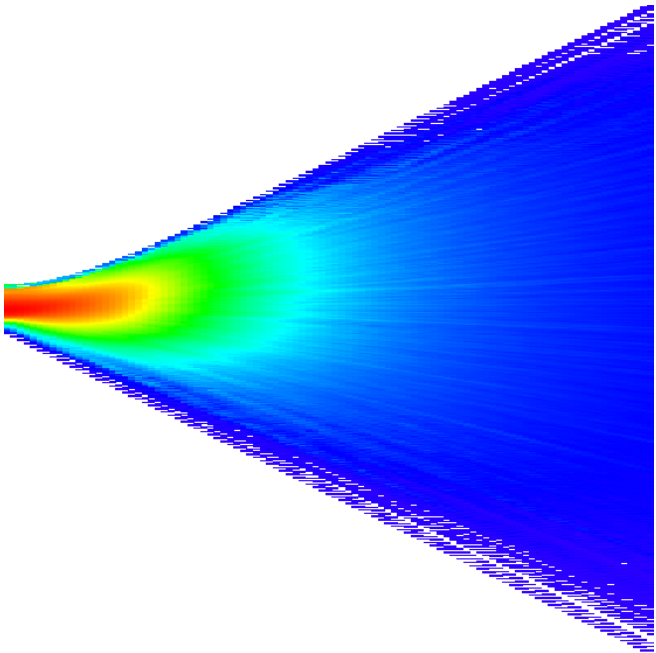}
\caption{(Color online) Entropy $\mathcal{S}_n$ as a function of
$\omega_g/d$, colors are proportional to  the entropy. Red (gray)
and blue (black) colors correspond to $\mathcal{S}_n \approx 13$ and
$\mathcal{S}_n = 0$ respectively; there are no eigenstates in the
region with white color. Horizontal axis gives the value of
$\omega_g/d$ from $\omega_g/d=0$ (left) to $\omega_g/d=20$ (right).
Vertical axis gives the energies of the central band eigenstates
(arbitrary units). The size of the lattice is $4\times 4$, $n_q=16$.
Disorder corresponds to $\delta=0.1$ and $\theta=0.3$rad. All the
points on the figure has been computed for the same realization of
the disorder.}
\end{figure}

In Fig.~3 we show the dependence of $\eta$ on the frequency
shift $\omega_g$ for different lattice sizes at a typical
values of angle $\theta=0.3$ and disorder strength $\delta=0.1$.
The transition from
a regular regime ($\omega_g>\omega_g^c$) to a chaotic one
($\omega_g<\omega_g^c$) is clearly seen for each presented lattice size.
In order to monitor the transition  more precisely
by the dependence of $\omega_g^c$ on parameters,
we define $\omega_g^c$ by the condition $\eta(\omega_g^c/d)=0.3$.
Fig.~3b shows $\eta$ as a function of
the rescaled frequency shift $\omega_g/\omega_g^c$.
The lower inset of Fig.~3b
clearly shows that $\omega_g^c$ is not a single-valued function of $n_q$.
This can be seen also in Fig.~3a where
the lattice configurations $6\times2$ and $4\times3$ corresponding to the
same number of qubits $n_q=12$ give two different curves of $\eta$ and
thus give two different values of $\omega_g^c$.
The upper inset of Fig.~3b shows that $\omega_g^c$ is
satisfactory described by a linear
dependence. A linear fit gives $\omega_g^c=1.26n_yd$.
Also the main panel of Fig.~3b shows that the transition
towards integrability becomes sharper as $n_y$ increases.
Hence, for a $100\times100$ array of $^{31}$P
spin-1/2  with a nearest-neighbour $^{31}$P-$^{31}$P distance
$b=4\mathrm{\AA}$, a magnetic field gradient of at least
$g_c=2.8$T$/\mu$m is needed in order to avoid emergence of quantum chaos
in the static quantum computer hardware.

The dependence found numerically
\begin{equation}\label{eq11}
\omega_g^c=C d n_y
\end{equation}
with a constant $C \approx 1.3$ can be understand on the basis of
the {\it {\AA}berg criterion} \cite{aberg}. This criterion has been
proposed \cite{aberg} to understand the conditions of emergence of
quantum chaos and ergodicity in many-body quantum systems. The
extensive numerical tests combined with analytical estimates have
been performed in
\cite{sushkov,Jacquod,georgeot00,georgeotshard,dls2001,benentiepjd,Berman1}
that allowed to establish the conditions of emergence of quantum
chaos and dynamical thermatization in variety of many-body quantum
systems. According to these results the quantum chaos emerges when
the coupling strength $U$ becomes comparable to the spacing between
directly coupled noninteracting states $\Delta_c$, \textit{i.e.}
when $U\sim\Delta_c$. It is important to stress that $\Delta_c$ is
exponentially larger than the level spacing between many-body
quantum states which for YQC is of the order of $\Delta_n \sim
\omega_gn_q 2^{-n_q}$.

\begin{figure}
\includegraphics[width=\columnwidth]{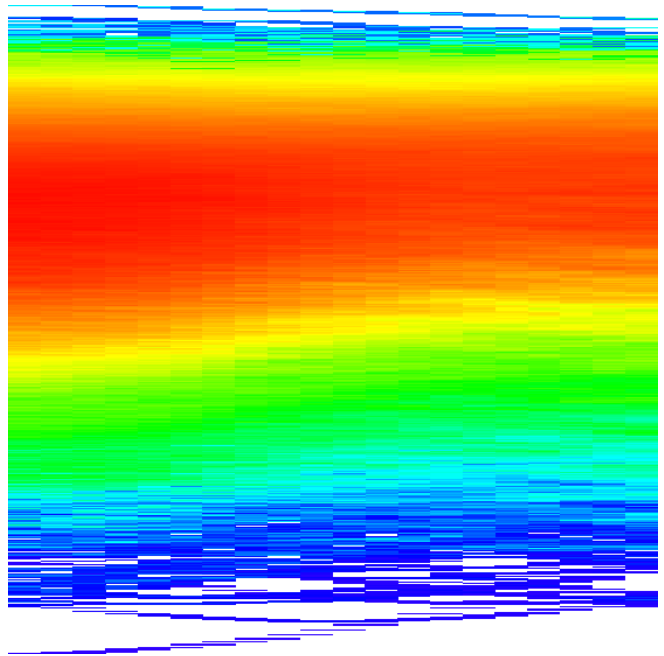}
\caption{(Color online) Entropy $\mathcal{S}_n$ for
$\omega_g/d=4$ as a function
of the magnetic field gradient angle $\theta$. Colors are as in Fig.~4.
Horizontal axis gives the value of $\theta$ from $\theta=0$ (left) to
$\theta=\pi/4$ (right). Vertical axis gives the energies of the
central band eigenstates (arbitrary units).
Disorder strength is $\delta=0.1$. All the points on the figure
has been computed for the same realization of the disorder as in Fig4
at the lattice size $4\times 4$.}
\end{figure}

In our model the noninteracting quantum register states are eigenstates
$\left\arrowvert\phi_i\right\rangle$ of the diagonal
Hamiltonian $H_{\mathrm{diag}}$ (\ref{eq9}). The off-diagonal part
$H_{\mathrm{off-diag}}$ (\ref{eq9}) contains flip-flop operators and
couples these noninteracting states thus being
responsible for the emergence of quantum chaos in the system.
The ensemble of the noninteracting states
$\left\{\left\arrowvert\phi_i\right\rangle\right\}_{i=1,N_H}$ forms
the quantum register basis, \textit{i.e.}, the basis is composed of
$N_H=2^{n_q}$ states with $N_H$-dimensional vectors written in
$\left(\left\arrowvert\uparrow\right\rangle,
\left\arrowvert\downarrow\right\rangle\right)$ or $\left(\left\arrowvert
0\right\rangle, \left\arrowvert 1\right\rangle\right)$ representation
(\textit{e.g.}
$\left\arrowvert\phi_i\right\rangle=\left\arrowvert0010100111
...01001\right\rangle$).
The quantum register  gives a  convenient computational basis
to perform quantum computations \cite{Nielsen},
each multiqubit state is then a linear
superposition of quantum register vectors $\left\arrowvert\phi_i\right\rangle$.

To apply the  {\AA}berg criterion to the YQC model we note that
the dipole interactions $d_{ij}=d/q_{ij}$
vanish quickly with the interqubit distance
and thus we can consider that only first
(at most second) nearest neighbours qubits are coupled together.
This means that $U \sim d$. For $\theta \ll 1$
the states coupled by transitions $H_{\mathrm{off-diag}}$ (\ref{eq9})
have a typical energy change $\omega_g$
(assuming that only few rows in $x$ contributes) and
the number of such transitions is of the order of $n_y$.
Thus $\Delta_c \sim \omega_g/n_y$ that leads to the result
(\ref{eq11}). For $\theta \sim 1$ still  the transitions
in the direction perpendicular to the field gradient
dominate so that in the relation (\ref{eq11})
$n_y$ should be replaced by $\sqrt{n_q}$ that gives
\begin{equation}\label{eq12}
\omega_g^c=C d \sqrt{n_q}  \;\; , (\theta \sim 1)
\end{equation}
with $C \sim 1$.

\begin{figure}
\includegraphics[width=\columnwidth]{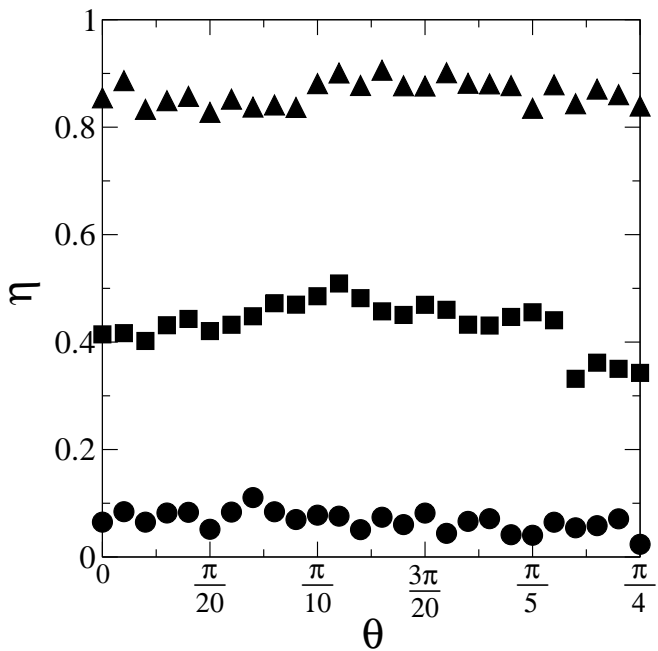}
\caption{Parameter $\eta$ as a function of the magnetic field
gradient angle $\theta$ for $\omega_g/d=3$ ($\bullet$), 4
($\blacksquare$), and 7 ($\blacktriangle$). The disorder strength is
$\delta=0.1$. The lattice size is $5\times 3$, $N_d=10$.}
\end{figure}

The emergence of quantum chaos manifests itself not only in the
level spacing statistics but also in the complexity
of quantum eigenstates in presence of interactions.
To demonstrate this fact in a quantitative way it is
convenient to define the complexity of an eigenvector
$\left\arrowvert\psi_n\right\rangle$ by its entropy
$\mathcal{S}_n=-\sum_{i=1}^{N_H}w_{ni}\log_2w_{ni}$ where $w_{ni}
=\left\arrowvert \left\langle \phi_i\arrowvert\psi_n \right\rangle
\right\arrowvert^2$. Here
$\left\{\left\arrowvert\phi_i\right\rangle\right\}_{i=1,N_H}$ is
the quantum register basis, or computational basis  composed of
$N_H=2^{n_q}$ states. Here, $N_H$-dimensional vectors are written in the
$\left(\left\arrowvert\uparrow\right\rangle,
\left\arrowvert\downarrow\right\rangle\right)$ or $\left(\left\arrowvert
0\right\rangle, \left\arrowvert 1\right\rangle\right)$ representation
(\textit{e.g.}
$\left\arrowvert\phi_i\right\rangle=\left\arrowvert0010100111
...01001\right\rangle$). The entropy $\mathcal{S}_n$ quantifies the
deviation of the eigenvector $\left\arrowvert\psi_n\right\rangle$
from a pure quantum register state. If $\mathcal{S}_n=0$,
$\left\arrowvert\psi_n\right\rangle=\left\arrowvert\phi_n\right\rangle$
is a quantum register state. If $\mathcal{S}_n=n_q$, all the quantum
register states $\left\arrowvert\phi_i\right\rangle$ are equally present
in the eigenvector $\left\arrowvert\psi_n\right\rangle$.
As we particularly focus on the central band eigenstates,
the maximum entropy attainable is
$\mathcal{S}_n = -\log_2
\binom{N}{\left[N/2\right]}<n_q$ for an eigenstate
$\left\arrowvert\psi_n\right\rangle$ equally composed
by all the quantum registers
 $\left\arrowvert\phi^B_i\right\rangle$
belonging to the $\left\langle I^z\right\rangle=0$ (even case) or
$\pm 1$ (odd case) sector. Indeed the symmetry of the Hamiltonian
(\ref{eq1}) and the fact that $\omega_0\gg\omega_g>d$ guarantee
that the central band eigenstates
$\left\arrowvert\psi_n\right\rangle$ can only be linear
superposition of the $\left\arrowvert\phi^B_i\right\rangle$
quantum registers.

Fig.~4 shows $\mathcal{S}_n$ as a function of $\omega_g/d$ for a
disorder strength $\delta=0.1$ and a magnetic field gradient angle
$\theta=0.3\mathrm{rad}$. It is clearly seen that the entropy of
eigenstates is reduced significantly with the increase of magnetic
field gradient. Thus a high gradient gives a suppression of quantum
chaos in the YQC model. The dependence of entropy on the angle
$\theta$ of magnetic field gradient is shown in Fig.~5. In agreement
with the theoretical estimates (\ref{eq11}), (\ref{eq12}) there is
no significant dependence on $\theta$. This is also true for the
level spacing statistics as it is demonstrated in Fig.~6 where the
parameter $\eta$ is clearly independent of $\theta$.

The dependence of $\eta$ on the disorder strength $\delta$ is
presented in Fig.~7. At weak and moderate values of rescaled field
gradient  $\omega_g/d$ an increase of $\delta$ drives the system to
a more integrable regime with higher values of $\eta$. This tendency
is similar to those found in \cite{georgeot00} where disorder also
stabilized the integrable phase. However, in the YQC model this
effect is weaker. Indeed, the increase of $\delta$ gives an increase
of $\Delta_c$ but not more than by a factor of two since there is
always a contribution from a regular field gradient. It is
interesting to note that according to the data of Fig.~7 in absence
of disorder the system is completely in the regime of quantum chaos
($\eta \approx 0$) at a small field gradient ($\omega_g/d \leq 2$).
This means that at $\omega_g/d \leq 2$ dipole-dipole interactions
between qubits lead to a melting of quantum computer hardware and
destruction of ideal computational basis.
\begin{figure}
\includegraphics[width=\columnwidth]{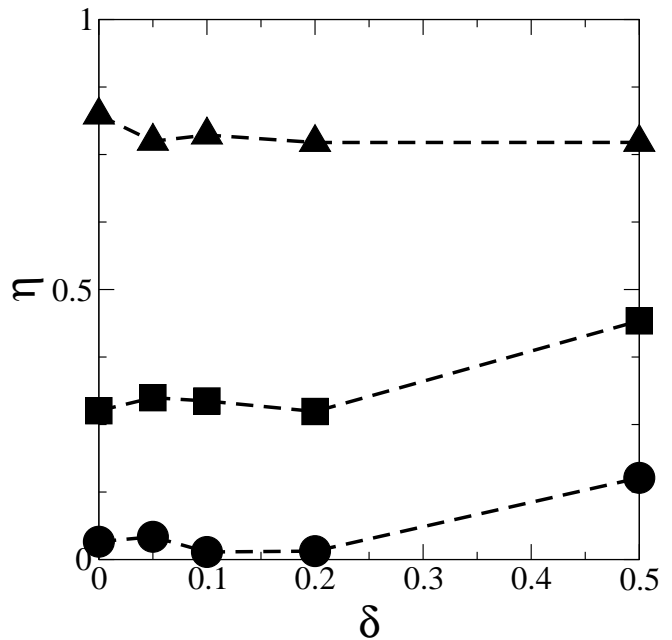}
\caption{Parameter $\eta$ as a function of the disorder strength
parameter $\delta$ for $\omega_g/d=2$ ($\bullet$), 4
($\blacksquare$), and 6 ($\blacktriangle$). The magnetic field
gradient angle is $\theta=0.3\mathrm{rad}$. The lattice size is
$5\times 3$, $N_d=10$.}
\end{figure}

\begin{figure}
\includegraphics[width=\columnwidth]{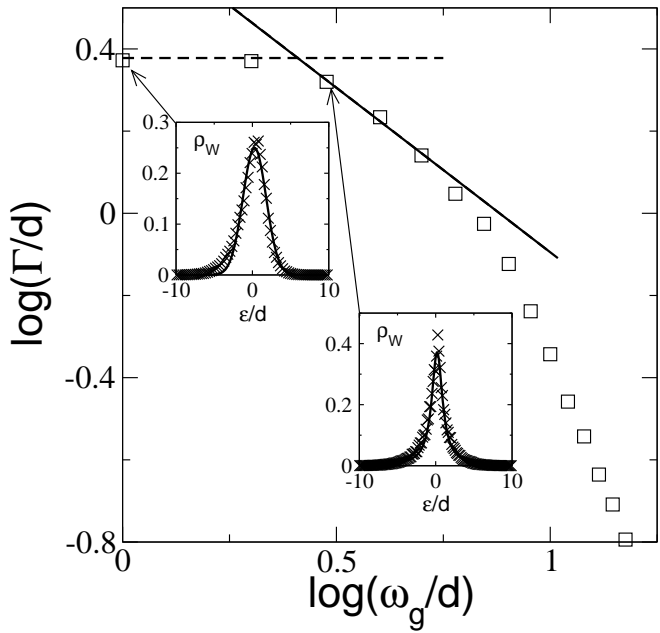}
\caption{Dependence of the width $\Gamma$ of the local density of
states $\rho_W$ on the frequency shift $\omega_g$ for a disorder
$\delta=0.1$ at a magnetic field gradient angle $\theta=0.3$rad and
a number of qubit $n_q=n_xn_y=5\times3$. The dashed line shows the
gaussian regime with $\Gamma\propto d$ and the full line shows the
Breit-Wigner regime with $\Gamma\propto d^2/\omega_g$. Lower inset:
an example of the local density of states $\rho_W$ for
$\omega_g/d=3$; the full curve shows the best fit of the
Breit-Wigner form with $\Gamma=2.1d$. Upper inset: example of the
local density of states $\rho_W$ for $\omega_g/d=1$; the full curve
shows the best Gaussian fit of width $\Gamma=2.52d$. In both insets
$\epsilon/d$ is rescaled energy, crosses give numerical data. The
transition towards integrability takes place near
$\log\left(\omega_g/d\right) \approx 0.7$ that corresponds to the
quantum chaos border (\ref{eq11}).}
\end{figure}

To determine the melting rate of the quantum computer hardware
we compute numerically the local density of states $\rho_W$
(see e.g. \cite{georgeot00}) defined as
\begin{equation}\label{eq13}
\rho_{W}(E-E_{i}) = \sum_{n} w_{ni}\delta(E-E_n)
\end{equation}
where $E_i$ is the energy of ideal quantum computational state
$\phi_i$ given by $H_{\mathrm{diag}}$ (\ref{eq8}). The energy width
$\Gamma$ of this distribution gives the rate on emergence of quantum
chaos \cite{georgeot00,flambaum}. Thus in the regime of quantum
chaos $\omega_g < \omega_g^c$ exponentially many states are mixed
after a time scale $1/\Gamma$ (here $\hbar=1$). The number of mixed
states is of the order of $\Gamma/\Delta_n$ where $\Delta_n$ is a
typical level spacing between many-body states. For weak couplings
between qubits the distribution $\rho_W(E)$ has the Breit-Wigner
shape with the width $\Gamma \sim U^2/\Delta_c$ (see e.g.
\cite{georgeot00,flambaum}). For the YQC case this gives
\begin{equation}\label{eq14}
\Gamma \sim d^2 \sqrt{n_q}/\omega_g \; .
\end{equation}
For strong couplings $d$ (at $\Gamma > \Delta_c$) the width starts
to grow linearly with coupling $\Gamma \sim d {n_q}^{1/4}$ similar
to the case considered in \cite{georgeot00,flambaum,flambaum1}. In
this case $\rho_W(E)$ has a gaussian shape. The numerical data shown
in Fig.~8 for 18 qubits confirm these theoretical formulas. Indeed,
for small $\omega_g/d$ the width $\Gamma$ is independent of
$\omega_g$ and $\rho_W(E)$ has a gaussian shape, while for larger
$\omega_g/d$ the distribution $\rho_W(E)$ starts to have the
Breit-Wigner shape and $\Gamma \propto 1/\omega_g$ in agreement with
(\ref{eq14}). For even larger values of $\omega_g/d$ one starts to
enter in the integrable regime $\omega_g > \omega_g^c$ where the
relation (\ref{eq14}) is not valid anymore. The place of the
transition is in a good agreement with the quantum chaos border
(\ref{eq11}) which gives $\omega_g^c = 3.9d$ (to be compared with
the approximate value $\omega_g^c \approx 5d$ obtained from data of
Fig.~8).

Thus an extensive amount of numerical data
obtained with up to 18 qubits give good agreement with the
theoretical estimates derived for the YQC model.

\section{Conclusion}

The numerical and analytical studies presented above establish the
parameter regime where the ideal computational basis of the quantum
computer model (YQC) proposed in \cite{Yamamoto} remains robust in
respect to dipole-dipole couplings between qubits. They clearly show
that a presence of magnetic field gradient allows to suppress
quantum chaos in the quantum computer hardware if the gradient $g$
is larger than the quantum chaos border given by Eqs. (\ref{eq11}),
(\ref{eq12}). For typical parameters of Table I (e.g. $^{29}$Si,
$b=1.9$\AA) the YQC hardware is in the stable regime at $g \approx
2$ T/$\mu$m for $n_q=100$ qubits and at $g \approx 20 T/\mu m$ for
$n_q=10^4$ qubits. These values can be realized with modern
experimental methods.

At the same time we should also note that
the stability of a quantum computer hardware
is not necessary sufficient for high accuracy
of quantum computations. Indeed, here we analyzed only
the static properties of YQC. In a realistic YQC implementation
it is also necessary to consider the accuracy of gate implementations
and the effects of static imperfections
on the accuracy of a concrete
operating quantum algorithm (see e.g. \cite{frahm}).
Future investigations are required to analyze
such operational accuracy of the YQC model \cite{Yamamoto}.

We thank T.D.Ladd and Y.Yamamoto
for discussions of the specific properties
of the quantum computer proposed in \cite{Yamamoto}.
This work was supported in part by the EC IST-FET project EDIQIP.

\end{document}